\documentclass[12pt,preprint]{aastex}
\newcommand{\etal}{{\it et al.}}
\newcommand{\teff}{$T_{{\rm eff}}$}

\shorttitle{The IMF in the Early Galaxy}
\shortauthors{Lucatello \etal}

\begin{document}


\title{Observational evidence for a different IMF in the early Galaxy}

\author{Sara Lucatello\altaffilmark{1,2}, Raffaele G. Gratton \altaffilmark{1},
Timothy C. Beers\altaffilmark{3}, Eugenio Carretta \altaffilmark{1,4}}
\altaffiltext{1}{INAF, Osservatorio Astronomico di Padova,  Vicolo dell'Osservatorio 5, 
        35122,  Padova,  Italy.}
\altaffiltext{2}{Dipartimento di Astronomia,  Universit\`a di Padova,  Vicolo dell'Osservatorio 2, 
        35122,  Padova,  Italy.}
\altaffiltext{3}{Department of Physics \& Astronomy and JINA: Joint Institute
for Nuclear Astrophysics,  Michigan State University, 
East Lansing,  Michigan 48824-1116.}
\altaffiltext{4}{INAF, Osservatorio Astronomico di Bologna, via Ranzani 1, 40127, Bologna, Italy} 

\begin{abstract}
The unexpected high incidence of carbon-enhanced, {\it s}-process enriched
unevolved stars amongst extremely metal-poor stars in the halo provides a
significant constraint on the Initial Mass Function (IMF) in the early Galaxy. We
argue that these objects are evidence for the past existence of a large
population of intermediate-mass stars, and conclude that the IMF in the early
Galaxy was different from the present, and shifted toward higher masses.
\end{abstract}


\keywords{stars: AGB and post-AGB --- stars: binaries --- Galaxy: stellar content --- Galaxy: fundamental parameters}


\section{Introduction}

The initial mass function (IMF) of stars has fundamental consequences
for the evolution of stellar systems, influencing the processes of metal
enrichment of the interstellar (and inter-galactic) medium, the expected
mass-to-light ratios of individual galaxies and globular clusters, and the
possible contribution of early-generation stars to re-ionization of the
universe. Hence, it is quite important to determine the form of the IMF over a
large range of physical conditions, in particular those typical of the very
early evolution of galaxies, when the metal content was still quite low (less
than 1\% of the solar value).

Determination of the early IMF from first principles is difficult, primarily due to
uncertainties related to the various processes involved in the fragmentation of
clouds and proto-stellar collapse (e.g., cooling, heating, magnetic fields,
turbulence, and rotation). A general expectation is that cooling is less
efficient at low metallicities, leading to preferential formation of more
massive objects (see, e.g., Bromm \& Larson 2004), but large uncertainties
still exist.

Observational determination of the IMF under conditions of very low metal
abundance is hampered by the fact that, in our galaxy, stars more massive than the
Sun that formed with [Fe/H] $< -2.0$ have already ended their nuclear burning
phases, and are now faint collapsed objects such as white dwarfs, neutron stars,
and black holes. Various authors (Ryu, Olive, \& Silk 1990; Adams \& Laughlin
1996, AL hereafter; Fields, Mathews, \& Schramm 1997) have argued for an IMF
that is peaked at intermediate masses for the metal-poor halo of our Galaxy.
These claims are based on constraints derived from the number density of white
dwarfs inferred from gravitational microlens experiments, coupled with those
given at low masses by the present halo luminosity, and at large masses by
nucleosynthesis limits. Such arguments are, however, still highly controversial
(Fields, Freese, \& Graff 2000; Gibson \& Mould 1997; M{\' e}ndez \& Minniti
2000; Majewski \& Siegel 2002).

Abia \etal (2001) set constraints on the early IMF on the basis of the large
fraction of carbon-rich stars that have been discovered by modern
surveys of extremely metal-poor (EMP) stars ({\it e.g.} Rossi, Beers, \& Sneden
1999; Norris, Beers \& Ryan 2000), and suggested that the IMF must have been
biased toward higher masses in the early Galaxy. However, we emphasize that this
result was based on then-incomplete knowledge of the statistics regarding the
fraction of carbon-enhanced EMP (CEMP) stars, and in particular concerning their
association with intermediate mass stars, hence the results obtained are highly
uncertain.
  
While direct observations of the early IMF are very difficult (if not
impossible) to obtain, one can search for evidence of their past existence with
presently observable stars. Such evidence can in particular be seen in binary
systems, where one of the components is a sub-solar-mass star, with a nuclear
evolutionary timescale comparable to the age of our Galaxy, and the other was an
intermediate-mass star (IMS; throughout this paper a star which undergoes the
third dredge-up during its Asymptotic Giant Branch--AGB--phase and becomes a 
C-star) that has gone
through all its nuclear phases, past its AGB stage, and is now a white dwarf.
During the AGB phase the IMS burns He into C in flashes; between flashes, a
restricted region within the star produces significant amounts of heavy elements
via the {\it s}-process (Straniero \etal{} 1995). AGB stars have extended outer
convective envelopes, and at the end of the flashes this convective envelope
penetrates into the region where nuclear burning previously occurred, dredging
up to the stellar surface large amounts of freshly produced carbon as well as
heavy nuclei produced through the {\it s}-process (Iben \& Renzini 1982). 

When an IMS evolves beyond the AGB stage, the carbon- and {\it s}-process-rich
envelope is lost to the interstellar medium through a slow wind. In a binary
system the companion may be able to capture part of this envelope (either
through direct mass transfer or accretion from the post-AGB wind of the IMS), and mix
the nucleosynthesis products with its outer layers. The presently observed
spectrum of the lower-mass companion of such a binary will still exhibit the
unmistakable chemical signature of AGB processing that occurred in the IMS
companion.

\section{Carbon-enhanced, {\it s}-process-Rich, Metal-Poor Stars and IMF Constraints}

The existence of carbon-rich binary systems among stars of solar composition is
long established (McConnell, Frye, \& Upgren 1972; Luck \& Bond, 1991). They are
called Barium stars at solar metallicities because of the presence of strong Ba
II lines, and CH stars at [Fe/H]$\sim -$1\ , due to the strong CH bands in
their spectra. These stars are also referred to as {\it extrinsic} carbon stars,
because they owe their large carbon abundances to pollution from a companion,
rather than to their own internal nucleosynthetic or mixing processes. About 1\%
of the stars in the Solar Neighborhood are classified as Ba or CH stars
(McConnell, Frye, \& Upgren 1972; Luck \& Bond 1991). 

The fraction of IMS in a population can hence be derived by the following simple formula, 
which takes into account the formation scenario of these objects:
\begin{equation}
f(IMS)=\frac{c(IMS)}{b\times p_{{\rm eff}}}.
\label{solmet}
\end{equation}
\noindent The term c(IMS) represents the fraction of stars which had IMS companions 
with the appropriate separations to transform them into the presently observed C- and {\it
s}-process enhanced stars, ({\it i.e.} a quantity directly inferred from observation). The
term $b$ is the binary fraction, which allows one to correct to the total population, and $p_{{\rm
eff}}$ indicates the percentage of binaries with orbital separation suitable for mass accretion to
take place.

We assume that about 60\% of stars with solar metallicities are in binary
systems (Jahrei{\ss} \& Wielen 2000), independent of metallicity. To estimate the range of initial
orbital periods for effective mass transfer, a few considerations are needed. 

The lower limit to the range of orbital separations is set by considering that
the stellar radius of the presumed donating companion should not exceed the
Roche-lobe radius during its previous evolutionary phases ({\it e.g.}, the red-giant branch).
This was derived adopting the Y2 isochrones (Yi, Kim,
\& Demarque 2003) at solar metallicity (period of $\sim$ 0.3 days). 

The upper limit of the component separations has been calculated by scaling that
adopted for metal-poor stars (upper limit to the orbital period of 250,000 days,
see below). In fact, the separation is expected to increase in inverse
proportion to the square root of the required enrichment; the latter increases
by a factor of $\sim 60$ with respect to stars of solar metallicity. 
The stars which we are considering have a typical metallicity of about
[Fe/H]$\simeq-$2.5, but are considered as carbon-rich stars only if their [C/Fe]
is at least a factor of ten higher than the solar ratio. This should be
contrasted with the factor of two enhancement that is often taken as a working
definition of carbon-rich stars amongst solar-metallicity stars. Therefore, the
upper limit to the period for solar metallicity stars is expected to be of order
$\sim$13,000\,days.  The fraction of binaries satisfying this assumption ({\it i.e.}
0.3$\leq$P$\leq$ 13,000) is estimated from the period distribution of Duquennoy \& Mayor 1991)
and is about $\sim$37\%. 
Substituting these values in Eq. \ref{solmet}, we obtain that
$\sim$5\% 
of solar-neighborhood stars were IMS. We assume that
the masses of the two components are independent of one another (Kroupa 1995); 
the adopted mass range for IMS is between 1.5 and 6\,M$_{\odot}$.
In fact, while at solar metallicity stars up to 8\,M$_{\odot}$ do undergo 
third dredge-up during their AGB phase, the objects in the upper end
of this range (6 to 8\,M$_{\odot}$) have quite effective hot bottom burning. 
C is burnt into N and thus the stars become N-rich rather than C-rich 
(see {\it e.g.} Forestini \& Charbonnel 1997, Marigo 2001 and Karakas 2003).
  Thus adopting a mass range between 1.5 and 6\,M$_{\odot}$ for the IMS and mass
cutoffs of 0.1 and 125\, M$_{\odot}$ for the lower and upper end of the stellar
masses, respectively, the Salpeter (1955) and Miller \& Scalo (1979; M\&S
hereafter) IMFs predict that, respectively, $\sim$2\% and $\sim$8\% of
Solar Neighborhood stars should be IMS. This quite good agreement found for
Population I suggests that we might derive the fraction of IMS in the total
stellar population (and hence obtain strong constraints on the IMF for all
stars) by simply counting the fraction of {\it extrinsic} carbon-enhanced stars
in a given sample. The rest of this paper exploits this method to constrain the
IMF of the old, metal-poor population of the Galactic halo. Of course, the first
challenge is to determine the fraction of the total population of CEMP stars that are
in fact extrinsic, rather than intrinsic. 

\section{The Importance of CEMP-s stars} 

The fraction of carbon-rich stars, among stars that have not yet reached
the AGB, is roughly constant down to a metallicity of a few percent of solar
(McConnell, Frye, \& Upgren 1972). However, a number of recent studies have
indicated that the fraction of carbon-rich stars increases abruptly amongst EMP
stars.

One of the most surprising results of the hitherto largest wide-field
spectroscopic survey for metal-poor stars, the HK survey (Beers, Preston \& Shectman
1992; Beers 1999), is the high frequency of carbon-enhanced stars found among
very metal-poor stars (Rossi, Beers \& Sneden 1999). This result has been confirmed by the Hamburg/ESO survey
(HES; Christlieb 2003), which found that $\sim$25-30\,\% of the metal-poor candidates
(with [Fe/H] $< -2.5$) appear to be carbon-enhanced stars. These unevolved CEMP
stars could have a variety of origins; Lucatello et al. (2004, Paper I) discuss at least
five possible classes of CEMP stars, the great majority of which, to date,
appear associated with {\it s}-process enrichment. Aoki et al. (2003) have
demonstrated that at least 70\% of a sample of 33 CEMP stars in their study exhibit
elements produced by the {\it s}-process, which we refer to as CEMP-s stars. As
discussed in Paper I, it is quite likely that {\it all} the
CEMP-s stars are members of binary systems.  The straightforward interpretation
is that 15-20\% of EMP stars had IMS companions with the proper
separations to transform them into presently-observed CEMP stars. 

\section{Derivation of the IMS fraction among EMP stars}

Taken at face value, the large fraction of CEMP-s stars amongst the CEMP objects
in the HK survey and HES suggests that a large number of IMS were present (as
companions in binary system) within this population. However, when comparing
this value with that obtained for more metal-rich stars, we must make a few
assumptions, and consider several possible selection effects that could bias the
determination. 

\subsection{Selection Effects}

{\bf Temperature scale:} Originally, both the HK survey and HES based their temperature scales on
$B-V$ colors. Estimates of the metal abundances were obtained by comparing the strength of the
Ca~II K line with that predicted for stars of the estimated $B-V$ color. However, it is well known
that the $B-V$ color is affected by the strength of the CH G-band (and other molecular carbon
features), leading to an underestimate of \teff{} in stars with strong bands. Therefore, a lower
metallicity was assigned to carbon-enhanced stars, with respect to carbon-normal stars with
identical atmospheric parameters. Hence, the parent population to be considered when comparing the
frequency of carbon-enhanced stars was larger, including stars more metal rich than the usual limit
for EMP stars. However, more recent results based on a considerable fraction of the HK survey data
adopt $V-K$ colors, which are more appropriate \teff{} indicators, as they are negligibly affected
by the strength of the CH G-band. Using these colors, it has been obtained that {\it at least} 20\%
of the stars with [Fe/H]$<-$2.5 are carbon-enhanced, {\it i.e.} with a carbon-to-iron ratio
obtained from intermediate-resolution spectroscopy of [C/Fe]$>+$1.0 (Beers et al. 2004, in
preparation). We adopt this value in our discussion.
 
{\bf Luminosity effect:} Carbon-enhancement, {\it per se}, does not affect the
bolometric luminosity of a given star. However, the presence of strong molecular
bands of CH and C$_{2}$ lowers the $B$ flux (and, to a smaller extent, the $V$
flux as well). Therefore, $B$- magnitude-limited surveys, such as the HK survey and the HES,
 will be biased against CEMP stars, given their lower luminosity in the
$B$ filter with respect to carbon-normal stars with the same abundance at the
same evolutionary stage. This effect leads to a small {\it underestimate} of the
CEMP fraction. In fact, the typical difference in $B$ magnitude is of order
$\sim$20\%; the volume sampled for carbon-enhanced stars will then be smaller by
$\sim$30\% than that considered for carbon-normal stars. We do not take this
effect into account in our analysis, but we remark that the adopted value is in
fact a conservative one.

{\bf Carbon-enhancement selection effect:} In order to be positively identified as carbon-enhanced
star, we have adopted the criterion that a given star must display an overabundance of
[C/Fe]$>$+1.0. Subgiants and RGB stars which are the surviving companions of a binary pair
including an IMS have already undergone first dredge-up, thus diluting the accreted material in
their convective envelopes. It is expected that such dilution could be as much as a factor of 20,
with the net effect of decreasing the observed atmospheric carbon abundance\footnote{This is
confirmed by the decrease of the Li abundance (see {\it e.g.,} Gratton \etal~2000; Spite et al.
2004)}. Therefore, many of the evolved stars which, in their main-sequence phases, had a carbon
overabundance well above the chosen threshold, dropped below this value after the first dredge-up
episode, and are thus identified as carbon-{\it normal} stars rather than carbon-enhanced ones,
affecting the measured fraction of the latter. Given their high luminosity, the evolved stars are
expected to make up the majority of the sample in these surveys, thus this effect could be very
important. In order to avoid this bias, the statistics should be limited exclusively to dwarf
stars. Indeed, recent results from the Sloan Digital Sky Survey, e.g. Downes \etal{} (2004), have
indicated that the number of dwarf carbon stars account about 60\% of all high-latitude C stars. If
it indeed proves to be the case that the dwarf carbon stars have their origin from mass transfer
from an IMS companion, which we suspect is likely, the appropriate fraction of IMS may be
substantially larger than the value adopted in the present discussion.

\subsection{Assumptions}

In making our calculations we are forced to adopt a number of assumptions, several of which are
still not well constrained. We discuss these assumptions in detail below.

{\bf Binary fraction:} The binary fraction for EMP stars is still not well known. However, for
slightly more metal-rich stars (e.g., [Fe/H] $\sim -2.0$) it is quite similar to that found for
stars of solar metal abundance (see {\it e.g.} Carney \etal{} 2003; Zapatero-Osorio \& Martin
2004). We will henceforth assume that the binary fraction is independent of metallicity, and adopt
the binary fraction of Jahrei{\ss} \& Wielen (2000), $\sim$60\%.

{\bf Separation range:} A metal-poor object is easily transformed into a
carbon-rich star, since, given its low Fe content, even a moderate amount of
processed matter accretion can produce the enhancement ([C/Fe]$>$+1) required to
identify an EMP star as CEMP. Therefore, the range of (original) separations
between components of a binary system that might be involved in the formation of
such objects is larger, as discussed in Paper I. The lower
limit on the range of component separations is set by considering that the
stellar radius of the presumed donating companion should not exceed the
Roche-lobe radius during its previous evolutionary phases (red-giant branch). We
adopted this limit from the the Y2 isochrones (Yi, Kim,
\& Demarque 2003) at [Fe/H]=$-$2.5.

For the present purpose, we estimate that the
range of initial orbital periods for CEMP-s stars ranges from $-$0.65$\leq \log
P\leq$5.4 ({\it i.e.} 0.2 to 250,000 days). Given the discussion in Paper I
, this is very likely an overestimate of the period range; however
this can be considered as a conservative estimate for such an interval. Adopting
the period distribution of Duquennoy \& Mayor (1991), this period range includes
59\% of all binaries. It is worth noting that, while this period range is not
well determined, its accuracy is not crucial for our conclusion. In fact, the
majority of binaries have separations within the useful range.

{\bf Mass range:} The mass threshold for stars to become carbon-rich when on the
AGB is likely to be smaller at lower metallicity. Fujimoto, Ikeda, \& Iben
(2000) argue that at metallicities below [Fe/H]$\sim-$3.5, stars with masses as
low as 0.8\,M$_{\odot}$ do indeed become carbon-enhanced in their AGB phase.
However, the typical metallicity for our objects is of about [Fe/H]=$-$2.5,
therefore we adopt 1.2\,M$_{\odot}$ as a more realistic lower limit for our
sample (Lattanzio 2003, private communication). 

The maximum mass for the third dredge-up during the AGB 
is expected to increase with decreasing
metallicity. However, the upper limit to the mass for becoming a C-star 
does decrease with metallicity, possibly reaching values as low as 
$\sim$3-4 for EMP stars
 (see Karakas 2003 and references therein). 
Keeping this in mind, we adopt as our definition of IMS at low metallicity 
stars with masses in the range 1.2 to 6\, M$_{\odot}$, noting that this is a 
{\it conservative} estimate. 
 
When all of the above factors are taken into account, substituting the adopted
values into Eq. \ref{solmet}, {\it i.e.}, $c(IMS)=14$\%, 
$b=60$\% and $p_{{\rm eff}}=59$\%,
 we find that the fraction of IMS companions amongst EMP stars is 40\%, 
a factor of eight larger than for solar-metallicity stars. Even if we
assume the most conservative scenario, wherein the binary fraction of EMP stars is 100\%,
 and demand that all IMS in binaries produce a carbon-rich companion
(irrespective of the initial system separation), the derived fraction of IMS
among EMPs is 14\%. The expected fraction of IMS, adopting a M\&S IMF is
$\sim$10\%, smaller than this (conservative) lower limit\footnote{If the lower
end of the IMS range is set to 0.9\,M$_{\odot}$, the M\&S IMF predicts the {\it
lower} limit of IMS derived from the data. However, models do not expect such
low values for the typical metallicity ([Fe/H]=$-$2.5) of our sample, see {\it
e.g.} Fujimoto, Ikeda \& Iben (2000)} . This result seems to favor an IMF peaked
at IMS for low-metallicity stars, as put forward by AL.  

Let us now assume for the IMF a log-normal form, 
\begin{equation}
\ln f (\ln m)=A-\frac{1}{2<\sigma>^{2}}[\ln (\frac{m}{m_{c}})]^{2}
\end{equation}
\noindent where $f=\frac{dN}{d\ln m}$. This general form for the IMF is motivated by star-
formation theory and by general statistical considerations (Larson 1973;
Elmegreen \& Mathieu 1983; Zinnecker 1984; Adams \& Fatuzzo 1996; Adams \&
Laughlin 1996). This form for the IMF is also sufficiently flexible to assume a
wide variety of behaviors. The parameter $A$ gives the overall normalization of
the distribution; $m_{c}$ is the mass scale, in solar masses, and sets the 
center of the distribution, while the dimensionless parameter $<\sigma>$
is the width of the distribution. Fitting such a function to our results, we
obtain values of 1.18 and 0.79\,M$_{\odot}$ respectively, for  $<\sigma>$ and
$m_{c}$. The corresponding values for the M\&S IMF are $<\sigma> \simeq$ 1.57 and
$m_{c}=$0.1\, M$_{\odot}$, while AL found, respectively, 0.44 and 2.3\,
M$_{\odot}$. 

\begin{figure}
\plotone{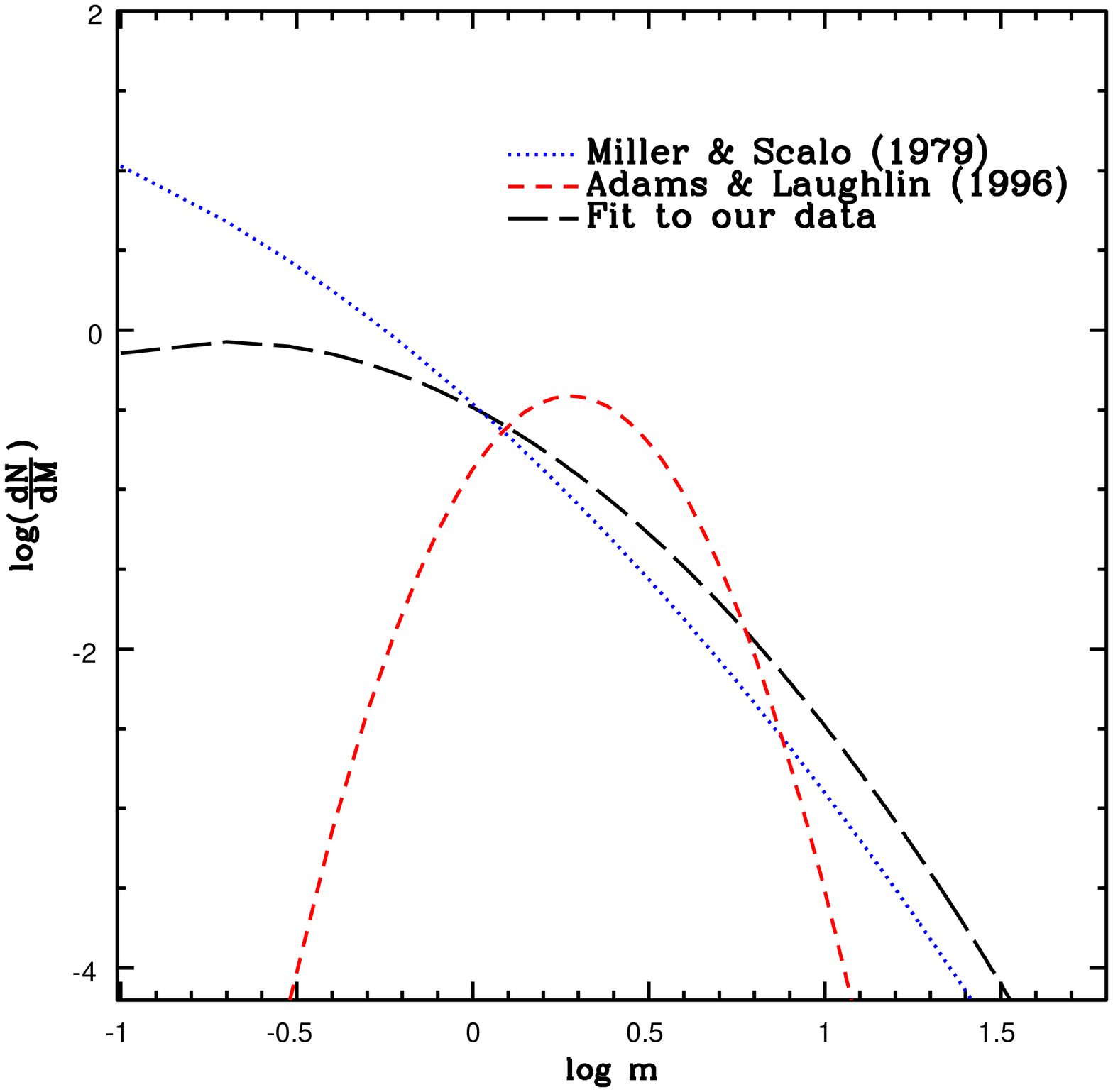}
\caption{Comparison of the IMF obtained assuming a log-normal form and fitting the
 measured IMS fraction (see text) with
those for M\&S and AL (1996).   
 \label{f_imf_imf}}
\end{figure}

As can be noted from inspection of Figure \ref{f_imf_imf}, which compares the
three above mentioned IMF's, as well as from the simple comparison of the curve
parameters, the IMF derived by fitting our results is not very different from
that of M\&S, showing basically a simple shift of the mass peak. On the other
hand, the shape of the AL IMF is peaked at much larger mass, and the mass range
for which the distribution function is not negligible is much narrower. We do
{\it not} claim that the IMF obtained by our fitting is an actual prediction of
the early IMF; in fact we do not have any direct information about high mass
stars. However, we point out that, in order to perform the fit we assumed the
same mass cutoffs as the M\&S IMF, {\it i.e.}, that the fraction of high-mass stars
formed is almost negligible.

\section{Discussion}

The primary objection against the existence of a large population of IMS in the
early Galaxy, including stars with metal abundances up to [Fe/H]=$-$2.0, is the
very large predicted production of carbon (Gibson \& Mould 1997; Fields, Freese,
\& Graff 2000). However, this objection does not apply to the EMP stars, which
comprise at most a few percent of the total halo population. In fact,
the EMP IMS contributed to the Galactic metal enrichment only at the end of their
nuclear-burning lives. Stars born with metallicities [Fe/H]$ \approx -3$ enrich the ISM
after a time ranging from 1.5 Gyr (1.5 M$_{\odot}$) down to 100 Myr (6
M$_{\odot}$) (Girardi \etal{} 1996).
 At  $\sim$10$^{8}$ years, the metal abundance of the interstellar medium is likely
to have been elevated to about [Fe/H]=$-$1.5 (Prantzos 2003), possibly reaching
a metal abundance as high as [Fe/H]=$-$1.0 (Chiappini, Romano, \& Matteucci 2003). 
A shift in the peak mass of the IMF also likely affects slightly larger mass stars, such as those 
those with masses ~6-8 Mo, which have shorter lifetimes
(a few 10$^{7}$ years for a metallicity of 
[Fe/H]$\simeq-$3, Girardi \etal{} 1996) and
 thus could in principle have an impact on the chemical composition of the other 
EMP stars. 
We do not have information about the mass distribution within the IMS range, 
however it is quite likely that the fraction of stars slightly more massive than IMS, 
although probably 
larger than among presently forming stars, was still very small and thus their effect on the 
ISM composition quite limited.

It is noteworthy that, the lower the metallicities, the shorter the nuclear burning lifetimes 
of IMS stars (see Chieffi \etal{} 2001), thus in principle, 
stars with metallicities lower than about [Fe/H]$\leq-$3.5 could contribute early on 
in the Galactic chemical evolution.  
Moreover, an early IMF biased toward larger masses could also have an impact on the 
white dwarf fraction, and thus on the SN Ia rate in the early Galaxy, which in turn affects
chemical evolution. 
However, when considering these effects,
 it is important to keep in mind that EMP stars account for only a very small fraction of 
the stellar content of the Galaxy, therefore a moderate shift of the early 
IMF towards higher masses should have a small impact on Galactic chemical evolution.
Hence, it is expected that:
\begin{list}{}{}
\item[i]
EMP IMS have likely not contributed significantly to the presently observed elemental 
compositions of other EMP stars, except for the outer envelopes of their companions in 
binary systems
\item[ii] The effect of EMP IMS on more metal-rich stars is diluted due to the 
contribution of the much larger population of moderately metal-poor ([Fe/H]$\sim -$1.5)
stars born from a ``conventional'' IMF
\end{list}   

In order to properly assess the effect of an IMF biased toward higher masses 
on the chemical evolution of the Galaxy, 
calculation of an accurate chemical evolution model would be necessary, which is beyond the
scope of the present paper.

An interesting consequence of the bias of the early IMF toward stars of higher
mass regards globular clusters (GC, hereafter). In fact, a GC forming with this
IMF would soon undergo large loss of gas from the system due to a population of
IMS. This gas could either result in the formation of new stars, be captured by
a pre-existing black hole in the cluster, or be lost to the ISM. The first two
hypotheses appear to be ruled out based on observations. There is no sign of
enrichment by {\it s}-process elements in GCs (with the exception of
$\omega$-Cen, which is, however a quite unusual case), and there is no evidence
for the presence of black holes in GCs with a mass comparable to that that
expected to be lost by the IMS population (see, {\it e.g.}, Gerssen \etal{} 2003).
Therefore, the only remaining possibility is that the material shed from the
IMSs is ejected into the ISM. This mass loss would have clear consequences on
the GC potential well, and eventually lead to the disruption of the cluster
itself. This could be the reason why there are no GCs at metallicities lower
than [Fe/H]=$-$2.5, and very few below $-$2\, dex.
%\footnote{On the other hand,
%GCs {\it do} exhibit huge variations of the abundances of light elements (N, O,
%Na) related to hot bottom burning (Gratton \etal{} 2001), but not to the
%triple-$\alpha$ reactions. This might be a signature of pollution by IMS more
%massive than those causing the CEMP-s phenomenon}. 

\section{Concluding remarks}

Our results indicate that the IMF of stars formed at early times in the
evolution of the Galaxy was different from that observed at present. In these
early epochs stars more massive than the Sun formed far more frequently than
now. {\it This is the first direct observational evidence that the IMF of stars
is not universal over time, lending support to the idea that the first
generations of stars included a substantial number of objects with very high
mass.}

This result, based on local samples of halo stars with well-measured elemental
abundances, provides stringent constraints on theoretical work aimed at deriving
the early IMF, as well as the interpretation of the abundance patterns in
distant objects such as those responsible for the Ly-$\alpha$ forest systems
(e.g., Songaila 2001; Pettini \etal 2003).

\acknowledgments
The authors thank John Norris and Chris Sneden
for useful discussions. S.L., R.G., and E.C. acknowledge
partial support from the MURST COFIN 2001. T.C.B. acknowledges partial funding
for this work from grants AST 00-98508 and AST 00-98549, as well as from grant
PHY 02-16783: Physics Frontiers Center/Joint Institute for Nuclear Astrophysics
(JINA), awarded by the U.S. National Science Foundation.

\clearpage

%\begin{figure}
%\plotone{logp_iso.ps}
%\caption{This is the first figure and it uses sgi9259.eps as
%its EPS figure file. \label{fig1}}
%\end{figure}
%
%\clearpage 
%
%\begin{figure}
%\plottwo{logp_lim.ps}{logp_lim.ps}
%\caption{This is an example of a multipart figure with a long figure caption 
%that must be set as a paragraph.  The processor has to buffer the text of the
%caption, so it is good not to be too wordy, but that would make for
%poor communication as well.\label{fig2}}
%\end{figure}

%% If you are not including electonic art with your submission, you may
%% mark up your captions using the \figcaption command. See the 
%% No more than seven \figcaption commands are allowed per page, 
%% so if you have more than seven captions, insert a \clearpage 

%% Tables should be submitted one per page, so put a \clearpage before

%% Two options are available to the author for producing tables:  the
%% deluxetable environment provided by the AASTeX package or the LaTeX
%% table environment.  Use of deluxetable is preferred.
%%

%% Three table samples follow, two marked up in the deluxetable environment,
%% one marked up as a LaTeX table.

%% In this first example, note that the \tabletypesize{}
%% command has been used to reduce the font size of the table.
%% Note also that the \label command needs to be placed 
\end{document}